\documentclass[conference]{IEEEtran}

\usepackage{cite}
\usepackage{amsmath,amssymb,amsfonts}
\usepackage{algorithmic}
\usepackage{graphicx}
\usepackage{mathtools}
\usepackage{color,soul}
\usepackage{epstopdf}

\begin{document}
\title{Fault Tolerance in Programmable Metasurfaces:\\ The Beam Steering Case}
\author{\IEEEauthorblockN{Hamidreza Taghvaee\IEEEauthorrefmark{1}, Sergi Abadal\IEEEauthorrefmark{1}, Julius Georgiou\IEEEauthorrefmark{2}, Albert Cabellos-Aparicio\IEEEauthorrefmark{1}, Eduard Alarc\'{o}n\IEEEauthorrefmark{1}}
\IEEEauthorblockA{\IEEEauthorrefmark{1}NaNoNetworking Center in Catalunya (N3Cat), Universitat Polit\`{e}cnica de Catalunya (UPC), Barcelona, Spain}
\IEEEauthorblockA{\IEEEauthorrefmark{2}Department of Electrical and Computer Engineering, University of Cyprus,
P.O. Box 20537, 1678, Nicosia, Cyprus}

}

\maketitle

\begin{abstract}
Metasurfaces, the two-dimensional counterpart of metamaterials, have caught great attention thanks to their powerful control over electromagnetic waves. Recent times have seen the emergence of a variety of metasurfaces exhibiting not only countless functionalities, but also a reconfigurable or even programmable response. Reconfigurability, however, entails the integration of tuning and control circuits within the metasurface structure and, as this new paradigm moves forward, new reliability challenges may arise. This paper examines, for the first time, the reliability problem in programmable metamaterials by proposing an error model and a general methodology for error analysis. To derive the error model, the causes and potential impact of faults are identified and discussed qualitatively. The methodology is presented and instantiated for beam steering, which constitutes a relevant example for programmable metasurfaces. Results show that performance degradation depends on the type of error and its spatial distribution and that, in beam steering, error rates over 10\% can still be considered acceptable.
\end{abstract}


\section{Introduction}
Metasurfaces are planar artificial structures composed by an array of subwavelength resonators that enable powerful control of electromagnetic waves \cite{Hsiao2017, Li2018, Chen2016, Tsilipakos2018a}. The resonator pattern modifies the impedance observed by the impinging wave, thus shaping the electromagnetic response of the metasurface through a particular transmittance or reflectance function. This way, metasurfaces implement planar lenses \cite{Nasari2014}, absorbers \cite{Taghvaee2014}, antennas \cite{Badawe2016, Hussain2017, Li2017c}, retroreflectors \cite{Arbabi2017}, optical mixers \cite{Liu2018a}, beam shaping \cite{Yoon2017}, or nonlinear devices \cite{Taghvaee2017, Tymchenko2017}.

\begin{figure}[!t]
\includegraphics[width=\columnwidth]{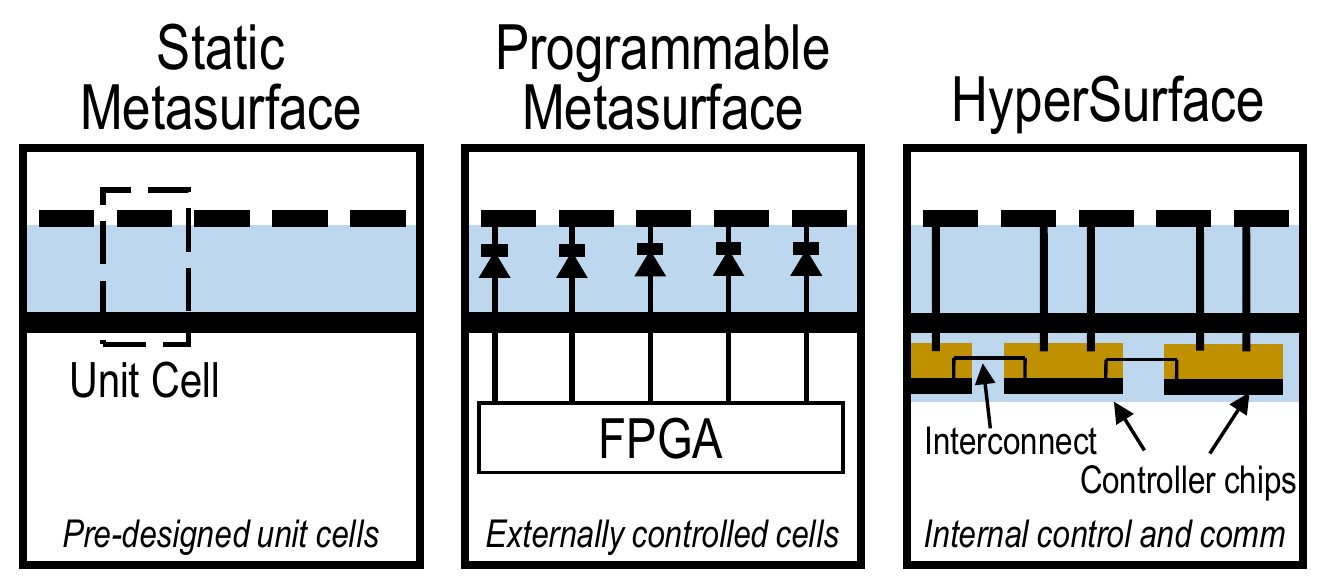} 
\vspace{-0.7cm}
\caption{Progression of metasurface design, from static to programmable with fully integrated tuning and control electronics.}
\vspace{-0.4cm}
\label{fig:intro}
\end{figure}

Metasurfaces are generally periodic and composed by a series of basic building blocks referred to as \emph{unit cells} (see Fig. \ref{fig:intro}), which are specifically designed for a given target electromagnetic purpose. This implies that, besides requiring specialists for its design, metasurfaces are typically static and non-reusable. Moreover, their resonant nature limits their functionalities, bandwidth, and operation conditions. Lately, however, reconfigurable metasurfaces have been proposed where the functionality can tuned or switched \cite{Oliveri2015, Zhang2017, Tsilipakos2018}, thereby improving applicability and adoption. Reconfigurability can be global or local depending on the actual means of tuning, including optical excitation \cite{Zhao2015} or electrostatic biasing \cite{Taghvaee2017a} or through the use of components such as diodes \cite{Yang2016a}, memristors \cite{Georgiou2018}, phase change materials \cite{Wang2016a} and microelectromechanical systems (MEMS) \cite{Han2015a, Kan2015}.

Recent works have taken one step further by combining local tunability with control methods, so that a single metasurface can be reprogrammed to fulfill different tasks \cite{Liu2018ISCAS, Cui2014, Liu2017a}. The example in \cite{Yang2016a} uses a Field-Programmable Gate Array (FPGA) to drive the pin diodes of a reconfigurable metasurface, where each pin diode determines the state of an individual unit cell. Beyond that, several authors have proposed to integrate a network of communicating chips within the metasurface containing actuators, control circuits, and even sensors \cite{AbadalACCESS, Tasolamprou2018, Liaskos2018, Liu2018}. This opens new opportunities in the design of autonomous self-adaptive programmable metasurfaces but, at the same time, poses significant challenges in the implementation, co-integration, and testing of the electronics within and around the metasurface.

This paper lays out the reliability problem in programmable metasurfaces and proposes a framework for its study. We build on the observation that metasurfaces will become prone to failure as they start integrating sophisticated tuning, control and sensing circuits (see Fig. \ref{fig:intro}). The design from \cite{Yang2016a} is a first example: the pin diodes enabling the reconfigurability may generate a considerable current density in the device, which could lead to electromigration. Further, works that consider the embedding of an array of controller chips within the metasurface structure \cite{Liaskos2018} will pose additional reliability concerns. On the one hand, technology is expected to be pushed seeking miniaturization and low power, but also making chips less reliable and fabrication mismatches more frequent \cite{Srinivasan2004}. On the other hand, metasurfaces may need to scale up for certain applications such as cloaking of a large object, e.g. a plane, thereby increasing the number of chips involved and introducing more points of failure. 

Despite all this, metasurfaces in general and programmable metasurfaces in particular have not been evaluated from the reliability perspective yet. This paper aims to bridge this gap by identifying potential causes of errors and analyzing their impact. To this end, as main contributions, we (i) discuss potential causes of error, (ii) derive a comprehensive error model that distinguishes between the spatial distribution and the individual effect of each fault, and (iii) outline a methodology for the error analysis. Although the proposed methodology is applicable to metasurfaces targeting any electromagnetic functionality, we instantiate it for a relevant use case, i.e. beam steering \cite{Wan2016, Huang2017, Tasolamprou2014, Tasolamprou2017}. By anticipating the impact of errors and identifying those that are most detrimental, our methodology could be applied to fault diagnosis or to provide guidelines for the design of reliable programmable metasurfaces \cite{Kouvaros2018}.


The remainder of this paper is as follows. Section \ref{sec:error} analyzes possible causes of errors and derives an error model. Section \ref{sec:methodology} presents the evaluation methodology, instantiated for the beam steering case. Section \ref{sec:results} shows the results of the analysis and Section \ref{sec:conclusions} concludes the paper.

\section{Error Model}
\label{sec:error}
To build a comprehensive error model, we focus on a prospective metasurface concept: the HyperSurface (HSF). Next, we first outline the architecture of a HSF in Sec. \ref{sec:struct} and discuss potential origins of faults in Sec. \ref{sec:origin}. This allows to derive an model that describes both the types of faults (Sec. \ref{sec:types}) and their expected spatial distribution (Sec. \ref{sec:spatial}).

\begin{figure}[!t]
\includegraphics[width=\columnwidth]{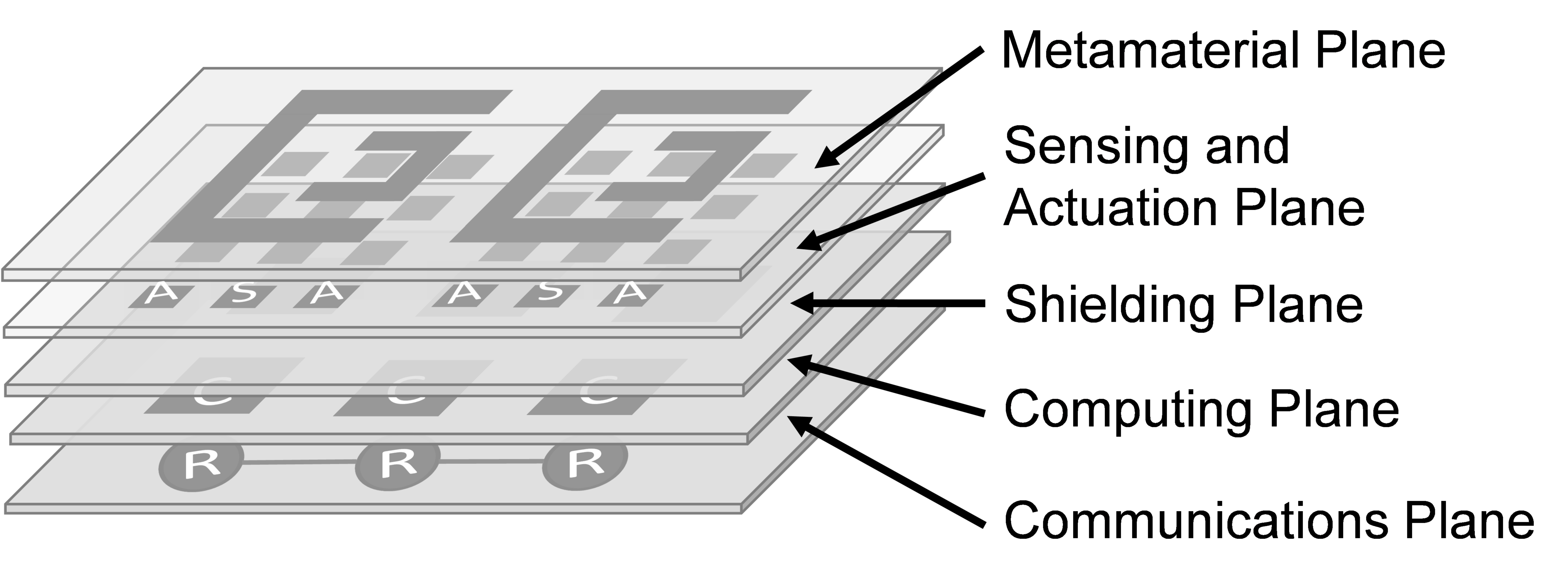} 
\vspace{-0.8cm}
\caption{Logical layered structure of a HyperSurface (HSF), which may include actuators (A), sensors (S), electronic controllers (C) and linked routers (R).}
\vspace{-0.4cm}
\label{fig:HSF}
\end{figure}

\subsection{The Hypersurface Structure}
\label{sec:struct}
A HSF is defined as a hardware platform whose electromagnetic behavior can be defined programmatically \cite{Liaskos2018}. In essence, a HSF consists on a reconfigurable metasurface with a network of sensors and controllers embedded in its structure. The intra-HSF network receives external programmatic commands and forwards them to the appropriate controllers, which alter the metasurface structure according to the desired electromagnetic behavior. Additionally, embedded sensors can help to maintain the desired state by adapting to changes in the environment without external intervention.

Figure \ref{fig:HSF} shows a graphical representation of the HSF structure, distinguishing the logic planes of the device. It seems clear that the addition of the computing plane (controller chips), the communication plane (network between controller chips), as well as the sensing and actuation plane, complicate the HSF design from the reliability perspective. Since each of these components can fail in several ways, we have chosen the HSF paradigm to guide our reasoning of the error model.

\subsection{Potential Origins of Faults}
\label{sec:origin}
Faults may occur for a wide variety of reasons that depend, among others, on the technology node, the manufacturing process, the HSF design, or the application environment. For instance, it is widely known that chip failure rates and fabrication mismatches increase as the technology nodes go down, therefore increasing the risk in more advanced designs targeting higher HSF frequencies with fine-grained control \cite{Srinivasan2004}. Manufacturing defects could lead to stuck unit cells, similar to dead pixels in displays. When interconnecting the chips that drive the different unit cells, connector constraints or bad fitting can also lead to errors of different typologies. 

Once deployed, chip connections to PCB might fail over time due to thermal cycling/flexing. Depending on the actual application environment, metasurface could be exposed to hazardous conditions that could lead to hard faults, such as physical damage in a conflict zones where bullets could impact the metasurface, or bit flips due cosmic radiation in space applications. Last but not least, ultra-low-power HSFs could power-gate a set of controllers in order to save energy in environments where a given performance degradation is tolerable. Here, the error analysis would help to determine which controllers should be powered off and at which state they should be kept. In any case, the power gating can be regarded as a intentional transient fault.


\subsection{Types of Errors}
\label{sec:types}
Let us model the metasurface as a matrix of unit cells, each of which is assigned a valid state $s \in \Sigma$ where $\Sigma$ is the set of valid states for a given metasurface design. The state $s$ basically determines the amplitude and phase of the impinging wave at an arbitrary unit cell. Then, we can distinguish between different types of errors, namely:
\begin{itemize}
\item \textbf{Stuck at state:} The unit cell is stuck at a random valid unit cell state, $s'\in \Sigma$. An error at the network or the controller may isolate the unit cell, which stays at a previous state forever. 
\item \textbf{Out of state:} The unit cell is stuck at a random invalid unit cell state, $s'\notin \Sigma$, which basically means random amplitude and phase. An error at the actuator or external biasing source leads to random levels or undefined switch states which means random response.
\item \textbf{Deterministic:} The unit cell stays in a known fixed value. A particularly relevant case of deterministic error would be a physically destroyed unit cell, which would be approximated as zero phase and full transmittance (zero reflection coefficient).
\item \textbf{Biased:} The unit cell is at a state which is at a fixed given distance $\Delta$ of the actual required state $s' = s+\Delta \in \Sigma$. This may be caused by flip-bit errors at the controller, or by external biases, perhaps due to by attacks. 
\end{itemize}

\begin{figure}[!t]
\centering
\includegraphics[width=0.7\columnwidth]{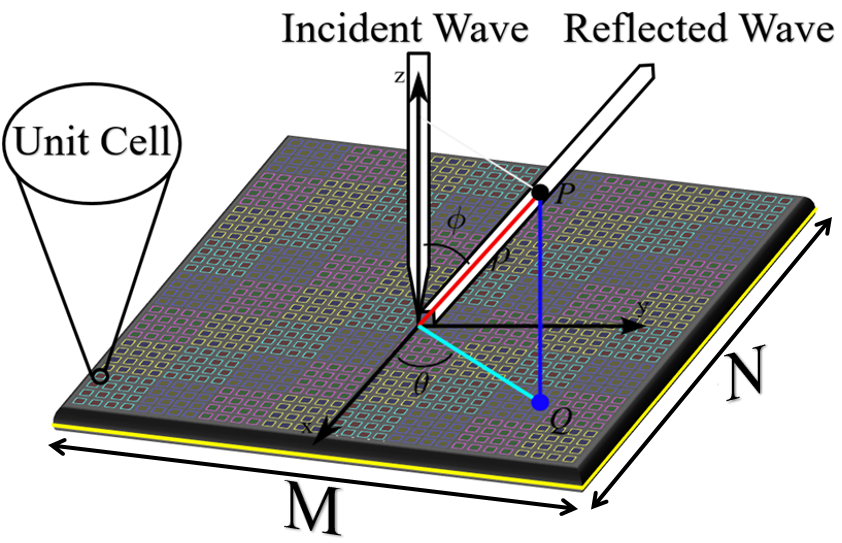}
\vspace{-.3cm}
\caption{Sketch of a reflecting metasurface with $N\times M$ unit cells and a normally incident plane wave.}
\label{fig:impinging}
\vspace{-0.4cm}
\end{figure}

\subsection{Spatial Distribution}
\label{sec:spatial}
As discussed earlier, some errors can show a spatial dependence due to the underlying cause. Due to this, we assume the following distributions:
\begin{itemize}
\item \textbf{Independent:} The errors are randomly distributed over the metasurface. Individual uncorrelated faults, maybe with different origins, could yield such a distribution.
\item \textbf{Clustered:} The errors appear mostly around a given area. In this case, cascading effects of a fault can lead to such behavior. Another possible source would be loss of connectivity at the network, leaving an entire region of the metasurface isolated and stuck in an old state. 
\item \textbf{Aligned:} Depending on the actual implementation of the metasurface, some actuators may be arranged in a regular fashion, perhaps distributing power or ground through a matrix of electrical lines. We speculate that, in such case, if one line representing a row or column fails, the whole row and column could be affected.
\item \textbf{State-specific:} Another speculative type of spatial distribution would be that all unit cells of a given region that are supposed to be in a specific state, behave incorrectly. This could happen if the actuator uses an external value (e.g. voltage from a centralized source) to determine given state; if that value is incorrect, the state will be delivered with a wrong value. 
\end{itemize}

\section{Methodology}
\label{sec:methodology}
A general methodology for the analysis of errors in metasurfaces would simply evaluate the metasurface in the presence of different error types and spatial distributions. To exemplify it, let us consider a certain metasurface design with $N\times M$ unit cells as shown in Fig. \ref{fig:impinging}. The metasurface implements a certain functionality (e.g., absorption, beam steering, polarization control) by setting each unit cell to a given state $s$. The performance of the metasurface is evaluated through the calculation of different metrics that depend on the actual functionality. The evaluation typically uses numerical methods in a full-wave electromagnetic solver, although such simulations are often resource and time consuming. Under certain conditions, the evaluation can be performed analytically instead. Although the methodology is general and can be applied to any electromagnetic functionality, we detail the steps followed in this paper to evaluate a beam steering reflecting metasurface.



\vspace{0.1cm}
\noindent
\textbf{Far-field calculation:} Beam steering is a particular case of wavefront manipulation that occurs in the far field. As such, the metasurface can be accurately modeled as an antenna array following the Huygens principle \cite{BalanisBOOK}. Therefore, considering each unit cell as an element of the array, the metasurface can be approached analytically and the far field is obtained as
\begin{equation}
F(\theta, \phi) = f_{E}(\theta, \phi) \cdot f_{A}(\theta, \phi) , 
\label{eq1}
\end{equation}
where $\theta$ is the elevation angle, $\phi$ is the azimuth angle of an arbitrary direction, $f_{E}(\theta, \phi)$ is the element factor (pattern function of unit cell) and $f_{A}(\theta, \phi)$ is the array factor (pattern function of unit cell arrangement). With the widespread assumptions that unit cells are isotropic and the excitation is a planar wave covering the entire metasurface, the scattering pattern will depend only on the array factor. For the metasurface shown in Fig. \ref{fig:impinging}, with $N\times M$ square unit cells, the far field pattern becomes
\begin{equation}
\begin{split}
&F(\theta, \phi) = \sum_{m=1}^{M} \sum_{n=1}^{N} A_{mn}e^{-j\zeta_{mn}} \\
&\zeta_{mn} = \Phi_{mn} + kL\sin{\theta}[(m-\tfrac{1}{2})\cos{\phi}+(n-\tfrac{1}{2}) \sin{\phi}]
\end{split}
\label{eq2}
\end{equation}
where $k$ is the wave number, $L$ is the lateral size the unit cell, whereas $A_{mn}$ and $\Phi_{mn}$ are the reflection amplitude and phase of the unit cell at position $(m,n)$, respectively. 

\vspace{0.1cm}
\noindent
\textbf{Introducing errors:} The analytical formulation allows to trivially introduce errors by modifying the terms $A_{mn}$ and $\Phi_{mn}$ of the affected unit cells. The type of error will determine the value of $A$ and $\Phi$, either random within a closed set of values (valid states) or random within a range (valid and invalid states), whereas the spatial distribution will affect $m$ and $n$. Simple algorithms are created to generate the required types and distributions with an arbitrary ratio of faults.





\vspace{0.1cm}
\noindent
\textbf{Beam steering metrics:} Conventional antenna theory can be used to evaluate beam steering. For the sake of brevity, we will only exemplify the methodology by calculating the directivity $D$ as
\begin{equation}
D(\theta, \phi)=10\log\frac{U(\theta, \phi)}{P_{tot}/4\pi}  \label{eq3}
\end{equation}
where $U(\theta, \phi)$ is the radiation intensity in a given direction, whereas $P_{tot}$ is the total radiated power. In our case, we can calculate the radiated power using the far-field pattern $F$ as
\begin{equation}
P_{tot}=\int_{0}^{2\pi}\int_{0}^{\pi}|F(\theta, \phi)|^2\sin(\theta, \phi)d\theta d\phi \label{eq4}.
\end{equation}
It is expected that, as errors are introduced, the performance metrics will decrease following a given trend. Identifying such tendencies is the main aim of this kind of analysis.

\section{Results}
\label{sec:results}
This section applies the proposed methodology on a particular beam steering metasurface. There are several ways to achieve beam steering, being gradient-index methods the most common \cite{Zhang2018a, Moccia2018, Xu2017b}. This means that unit cells generate a metasurface-wide phase gradient, whose value depends on the target direction. A few unit cell states are required to attain reasonable performance and, therefore, we will assume a metasurface design with four possible states with $A = 1$ and $p={0, \pi/2, \pi, 3\pi/2}$. We consider $M=N=60$ and design the metasurface to point to $\theta=\phi=\pi/4$ in all cases. 

\begin{figure}[!t]
\centering
\includegraphics[width=0.9\columnwidth]{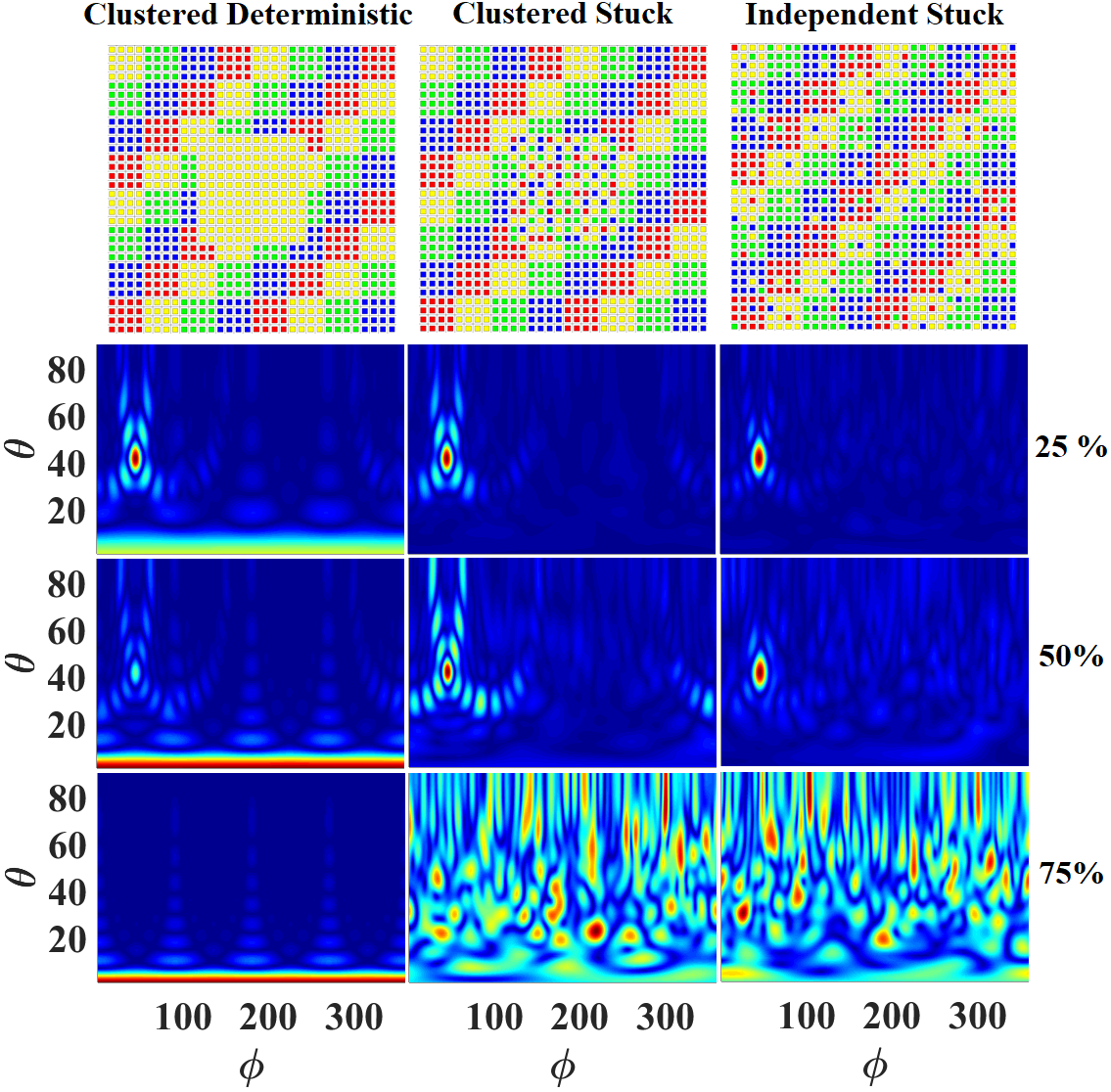}
\vspace{-0.3cm}
\caption{Sketch of the unit cell states and reflecting patterns for different error percentages (rows) and different error combinations (columns).}
\label{mixed}
\vspace{-0.45cm}
\end{figure}

Figure \ref{mixed} demonstrates how different types of error and their spatial distribution can behave very differently. Each column in the figure illustrates a different combination of error type and spatial distribution, whereas the top row shows the the unit cell states (each color is a given state) and the rest of rows plot the reflection pattern of the metasurface for different error percentages. The metasurface points most of the energy towards $\theta=\phi=\pi/4$ for relatively low error percentages and starts losing its functionality as the percentage increases. The differences among the distinct types of errors are clearly distinguishable. For instance, a clustered deterministic error in the center (e.g. an undesired partial reset) turns the beam steering metasurface into a specular reflector, as all the energy is reflected normally to the metasurface. In the clustered stuck case, side beams start appearing early and end up concealing the desired beam. Finally, it is observed how the independent stuck case leads to a rather uniform redistribution of the energy, less harmful at low percentages. From these plots, a partial conclusion would be that clustered deterministic errors would be the most detrimental, whereas uncorrelated and random errors would be most tolerable. 

\begin{figure}[!t]
\centering
\includegraphics[width=0.75\columnwidth]{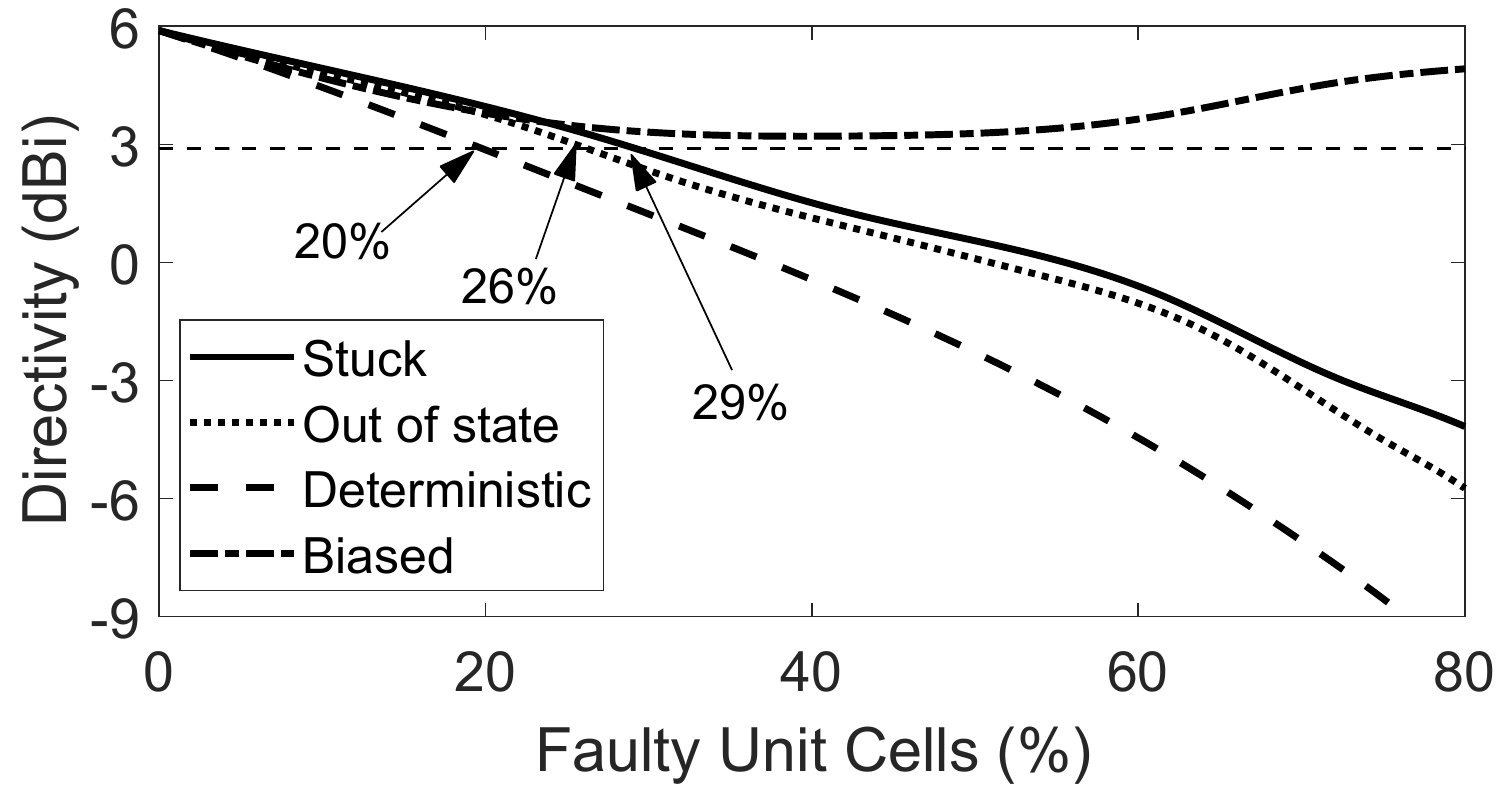}
\vspace{-0.4cm}
\caption{Directivity of the beam steering metasurface as a function of the error percentage for different error types.}
\label{types}
\vspace{-0.15cm}
\end{figure}

To fairly conclude about type and spatial distribution of errors, they need to be investigated separately. Figure \ref{types} illustrates the impact of the different types of errors by plotting the directivity over error percentage (spatial distribution is set to independent). The insets indicate the percentage of errors that lead to a loss of 3 dB in directivity. As expected, the most detrimental type is deterministic because all wrong values are mapped to same phase, which has more detrimental effect in the beam steering case due to its phase-gradient requirements. This reasoning implies that different types of errors may have a completely different impact on metasurfaces implementing different functionalities: for instance, absorbers may set the same value to all unit cells and, therefore, deterministic errors may not reduce performance. Such a behavior is actually observed in Figure \ref{types} as well, where it is shown that biased errors degrade performance up to a certain point and then return to nominal performance. This is because shifting the state of all the unit cells does not change the phase gradient.

\begin{figure}[!t]
\centering
\includegraphics[width=0.75\columnwidth]{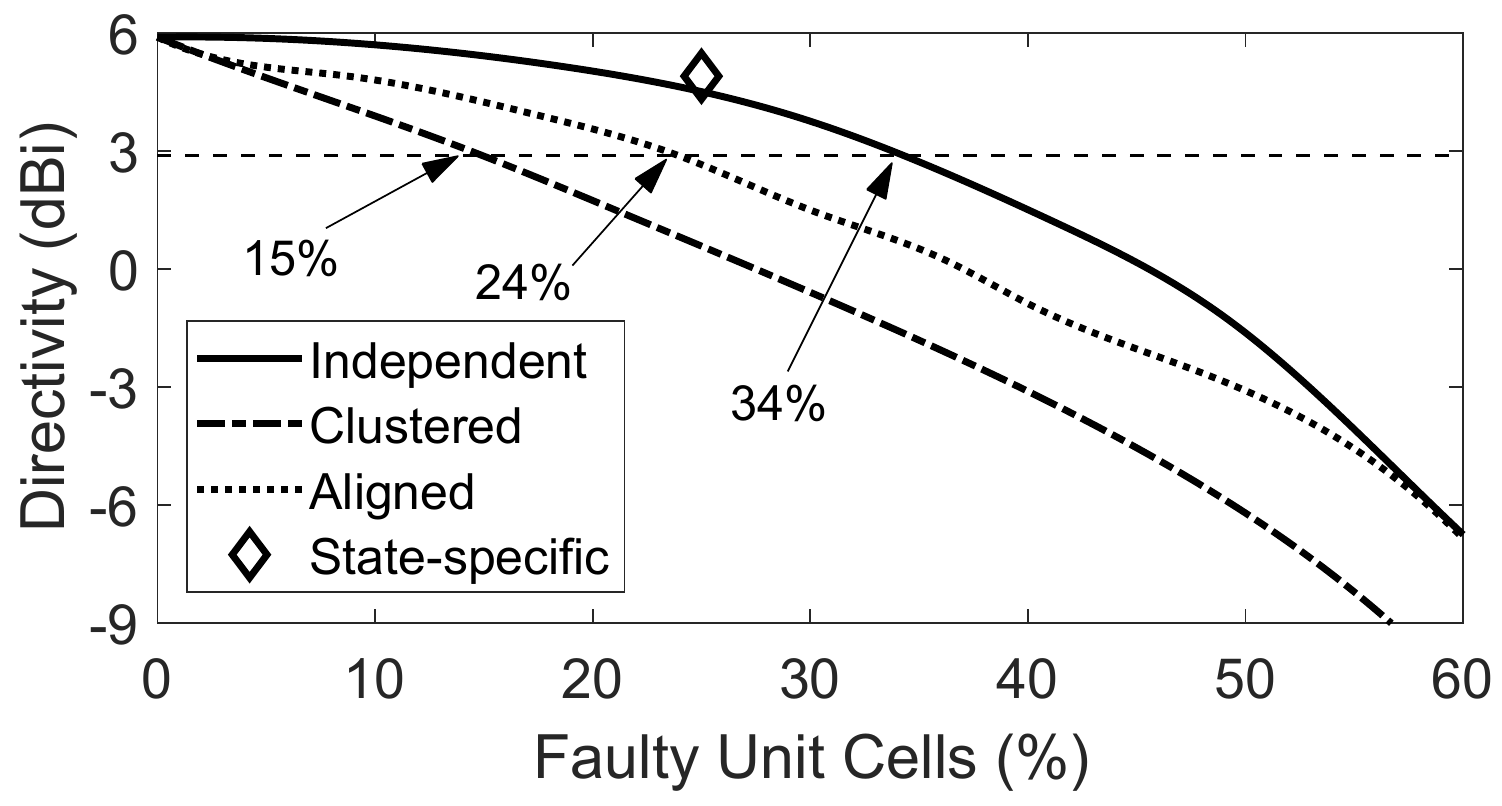}
\vspace{-0.4cm}
\caption{Directivity of the beam steering metasurface as a function of the error percentage for different error spatial distributions.}
\label{Distribution}
\vspace{-0.15cm}
\end{figure}

Finally, Figure \ref{Distribution} plots the directivity as a function of the error percentage for different spatial distributions (type of error is set to stuck at state). It is observed that the clustered distribution is the worst case for the reasons outlined above, whereas independent error is the most tolerable. The difference between both to reach the 3 dB loss point is a significant 19\%. In case of state-specific error, we can only report one single data representing 25\% error, as that is the percentage of unit cells set to phase 0 and that we considered erroneous. This value might change depending on the actual functionality.

\section{Conclusion}
\label{sec:conclusions}
We have proposed an error model and a methodology for error analysis in metasurfaces. Beam steering metasurfaces, where the functionality depends on the phase gradient, are robust against spatially uncorrelated errors with random values and to attacks that only bias the state of the unit cells. On the contrary, clustered errors that set all the unit cells to the same state are very detrimental. These results show the value of the error analysis and suggest that the error model is comprehensive enough to cover all possible cases. Future works will further analyze the impact of errors in metasurfaces with different sizes or functionalities, to then derive useful design guidelines and power-gating directives.

%

\section*{Acknowledgment}
This work has been supported by the European Commission under grant H2020-FETOPEN-736876 (VISORSURF) and by ICREA under the ICREA Academia programme.



\begin{thebibliography}{10}
\providecommand{\url}[1]{#1}
\csname url@samestyle\endcsname
\providecommand{\newblock}{\relax}
\providecommand{\bibinfo}[2]{#2}
\providecommand{\BIBentrySTDinterwordspacing}{\spaceskip=0pt\relax}
\providecommand{\BIBentryALTinterwordstretchfactor}{4}
\providecommand{\BIBentryALTinterwordspacing}{\spaceskip=\fontdimen2\font plus
\BIBentryALTinterwordstretchfactor\fontdimen3\font minus
  \fontdimen4\font\relax}
\providecommand{\BIBforeignlanguage}[2]{{%
\expandafter\ifx\csname l@#1\endcsname\relax
\typeout{** WARNING: IEEEtran.bst: No hyphenation pattern has been}%
\typeout{** loaded for the language `#1'. Using the pattern for}%
\typeout{** the default language instead.}%
\else
\language=\csname l@#1\endcsname
\fi
#2}}
\providecommand{\BIBdecl}{\relax}
\BIBdecl

\bibitem{Hsiao2017}
H.-H. Hsiao, C.~H. Chu, and D.~P. Tsai, ``{Fundamentals and Applications of
  Metasurfaces},'' pp. 1--20, 2017.

\bibitem{Li2018}
A.~Li, S.~Singh, and D.~Sievenpiper, ``{Metasurfaces and their applications},''
  \emph{Nanophotonics}, vol.~7, no.~6, pp. 989--1011, 2018.

\bibitem{Chen2016}
H.~T. Chen, A.~J. Taylor, and N.~Yu, ``{A review of metasurfaces: Physics and
  applications},'' \emph{Reports on Progress in Physics}, vol.~79, no.~7, 2016.

\bibitem{Tsilipakos2018a}
O.~Tsilipakos, A.~C. Tasolamprou, T.~Koschny, M.~Kafesaki, E.~N. Economou, and
  C.~M. Soukoulis, ``{Pairing Toroidal and Magnetic Dipole Resonances in
  Elliptic Dielectric Rod Metasurfaces for Reconfigurable Wavefront
  Manipulation in Reflection},'' \emph{Advanced Optical Materials}, vol.~6,
  no.~22, 2018.

\bibitem{Nasari2014}
H.~Nasari and M.~S. Abrishamian, ``{Electrically tunable graded index planar
  lens based on graphene},'' \emph{Journal of Applied Physics}, vol. 116,
  no.~8, 2014.

\bibitem{Taghvaee2014}
H.~Taghvaee, M.~Seyyedi, and A.~Rezaee, ``{Design of a Metamaterial Dual Band
  Absorber},'' in \emph{The third Iranian Conference on Engineering
  Electromagnetic ICEEM}, vol.~3.\hskip 1em plus 0.5em minus 0.4em\relax
  Tehran: Iranian Scientific Society of Engineering Electromagnetics, 2014.

\bibitem{Badawe2016}
M.~E. Badawe, T.~Almoneef, and O.~M. Ramahi, ``{A true metasurface antenna},''
  \emph{2016 IEEE Antennas and Propagation Society International Symposium,
  APSURSI 2016 - Proceedings}, no. January, pp. 1903--1904, 2016.

\bibitem{Hussain2017}
N.~Hussain and I.~Park, ``{Design of a wide-gain-bandwidth metasurface antenna
  at terahertz frequency},'' \emph{AIP Advances}, vol.~7, no.~5, 2017.

\bibitem{Li2017c}
X.~Li, J.~Yang, Y.~Feng, M.~Yang, and M.~Huang, ``{Compact and broadband
  antenna based on a step-shaped metasurface},'' \emph{Optics Express},
  vol.~25, no.~16, p. 19023, 2017.

\bibitem{Arbabi2017}
A.~Arbabi, E.~Arbabi, Y.~Horie, S.~M. Kamali, and A.~Faraon, ``{Planar
  metasurface retroreflector},'' \emph{Nature Photonics}, vol.~11, no.~7, pp.
  415--420, 2017.

\bibitem{Liu2018a}
S.~Liu, P.~P. Vabishchevich, A.~Vaskin, J.~L. Reno, G.~A. Keeler, M.~B.
  Sinclair, I.~Staude, and I.~Brener, ``{An all-dielectric metasurface as a
  broadband optical frequency mixer},'' \emph{Nature Communications}, vol.~9,
  no.~1, pp. 1--6, 2018.

\bibitem{Yoon2017}
I.~Yoon and J.~Oh, ``{Millimeter wave thin metasurface enabling
  polarization-controlled beam shaping},'' \emph{2017 IEEE Antennas and
  Propagation Society International Symposium, Proceedings}, vol. 2017-Janua,
  pp. 669--670, 2017.

\bibitem{Taghvaee2017}
H.~R. Taghvaee, F.~Zarrinkhat, and M.~S. Abrishamian, ``{Terahertz Kerr
  nonlinearity analysis of a microribbon graphene array using the harmonic
  balance method},'' \emph{Journal of Physics D: Applied Physics}, vol.~50,
  no.~25, 2017.

\bibitem{Tymchenko2017}
M.~Tymchenko, J.~S. Gomez-Diaz, J.~Lee, M.~A. Belkin, and A.~Alu,
  ``{Highly-efficient THz generation using nonlinear plasmonic metasurfaces},''
  \emph{Journal of Optics}, vol.~19, no.~10, p. 104001, 2017.

\bibitem{Oliveri2015}
G.~Oliveri, D.~Werner, and A.~Massa, ``{Reconfigurable electromagnetics through
  metamaterials - A Review},'' \emph{Proceedings of the IEEE}, vol. 103, no.~7,
  pp. 1034--1056, 2015.

\bibitem{Zhang2017}
M.~Zhang, W.~Zhang, A.~Q. Liu, F.~C. Li, and C.~F. Lan, ``{Tunable Polarization
  Conversion and Rotation based on a Reconfigurable Metasurface},''
  \emph{Scientific Reports}, vol.~7, no.~1, pp. 1--7, 2017.

\bibitem{Tsilipakos2018}
O.~Tsilipakos, F.~Liu, A.~Pitilakis, A.~C. Tasolamprou, D.-H. Kwon, M.~S.
  Mirmoosa, N.~Kantartzis, E.~N. Economou, M.~Kafesaki, C.~M. Soukoulis, and
  S.~A. Tretyakov, ``{Tunable perfect anomalous reflection in metasurfaces with
  capacitive lumped elements},'' in \emph{Proceedings of METAMATERIALS '18},
  2018, pp. 392--394.

\bibitem{Zhao2015}
X.~Zhao, K.~Fan, J.~Zhang, H.~R. Seren, G.~D. Metcalfe, M.~Wraback, R.~D.
  Averitt, and X.~Zhang, ``{Optically tunable metamaterial perfect absorber on
  highly flexible substrate},'' \emph{Sensors and Actuators, A: Physical}, vol.
  231, pp. 74--80, 2015.

\bibitem{Taghvaee2017a}
H.~R. Taghvaee, H.~Nasari, and M.~S. Abrishamian, ``{Circuit modeling of
  graphene absorber in terahertz band},'' \emph{Optics Communications}, vol.
  383, pp. 11--16, 2017.

\bibitem{Yang2016a}
H.~Yang, X.~Cao, F.~Yang, J.~Gao, S.~Xu, and M.~Li, ``{A programmable
  metasurface with dynamic polarization , scattering and focusing control},''
  \emph{Scientific Reports}, no. October, pp. 1--11, 2016.

\bibitem{Georgiou2018}
J.~Georgiou, K.~M. Kossifos, M.~A. Antoniades, A.~H. Jaafar, and {N. T. Kemp},
  ``{Chua Mem-Components for Adaptive RF Metamaterials},'' in \emph{Proceedings
  of the ISCAS '18}, 2018.

\bibitem{Wang2016a}
D.~Wang, L.~Zhang, Y.~Gong, L.~Jian, T.~Venkatesan, C.~W. Qiu, and M.~Hong,
  ``{Multiband switchable terahertz quarter-wave plates via phase-change
  metasurfaces},'' \emph{IEEE Photonics Journal}, vol.~8, no.~1, 2016.

\bibitem{Han2015a}
Z.~Han, K.~Kohno, H.~Fujita, K.~Hirakawa, and H.~Toshiyoshi, ``{Tunable
  terahertz filter and modulator based on electrostatic MEMS reconfigurable SRR
  array},'' \emph{IEEE Journal of Selected Topics in Quantum Electronics},
  vol.~21, no.~4, 2015.

\bibitem{Kan2015}
T.~Kan, A.~Isozaki, N.~Kanda, N.~Nemoto, K.~Konishi, H.~Takahashi,
  M.~Kuwata-Gonokami, K.~Matsumoto, and I.~Shimoyama, ``{Enantiomeric switching
  of chiral metamaterial for terahertz polarization modulation employing
  vertically deformable MEMS spirals},'' \emph{Nature Communications}, vol.~6,
  pp. 1--7, 2015.

\bibitem{Liu2018ISCAS}
F.~Liu, A.~Pitilakis, M.~S. Mirmoosa, O.~Tsilipakos, X.~Wang, A.~C.
  Tasolamprou, S.~Abadal, A.~Cabellos-Aparicio, E.~Alarc{\'{o}}n, C.~Liaskos,
  N.~V. Kantartzis, M.~Kafesaki, E.~N. Economou, C.~M. Soukoulis, and
  S.~Tretyakov, ``{Programmable Metasurfaces: State of the art and
  Prospects},'' in \emph{Proceedings of the ISCAS '18}, 2018.

\bibitem{Cui2014}
T.~J. Cui, M.~Q. Qi, X.~Wan, J.~Zhao, Q.~Cheng, K.~T. Lee, J.~Y. Lee, S.~Seo,
  L.~J. Guo, Z.~Zhang, Z.~You, and D.~Chu, ``{Coding metamaterials, digital
  metamaterials and programmable metamaterials},'' \emph{Light: Science and
  Applications}, vol.~3, no.~10, pp. 1--9, 2014.

\bibitem{Liu2017a}
S.~Liu and T.~J. Cui, ``{Flexible Controls of Terahertz Waves Using Coding and
  Programmable Metasurfaces},'' \emph{IEEE Journal of Selected Topics in
  Quantum Electronics}, vol.~23, no.~4, 2017.

\bibitem{AbadalACCESS}
S.~Abadal, C.~Liaskos, A.~Tsioliaridou, S.~Ioannidis, A.~Pitsillides,
  J.~Sol{\'{e}}-Pareta, E.~Alarc{\'{o}}n, and A.~Cabellos-Aparicio,
  ``{Computing and Communications for the Software-Defined Metamaterial
  Paradigm: A Context Analysis},'' \emph{IEEE Access}, vol.~5, pp. 6225--6235,
  2017.

\bibitem{Tasolamprou2018}
A.~C. Tasolamprou, M.~S. Mirmoosa, O.~Tsilipakos, A.~Pitilakis, F.~Liu,
  S.~Abadal, A.~Cabellos-Aparicio, E.~Alarc{\'{o}}n, C.~Liaskos, N.~V.
  Kantartzis, S.~Tretyakov, M.~Kafesaki, E.~N. Economou, and C.~M. Soukoulis,
  ``{Intercell wireless communication in software-defined metasurfaces},'' in
  \emph{Proceedings of the ISCAS '18}, 2018.

\bibitem{Liaskos2018}
C.~Liaskos, A.~Tsioliaridou, A.~Pitsillides, S.~Ioannidis, and I.~Akyildiz,
  ``{Using any Surface to Realize a New Paradigm for Wireless
  Communications},'' 2018.

\bibitem{Liu2018}
\BIBentryALTinterwordspacing
F.~Liu, O.~Tsilipakos, A.~Pitilakis, A.~C. Tasolamprou, M.~S. Mirmoosa, N.~V.
  Kantartzis, D.-h. Kwon, M.~Kafesaki, C.~M. Soukoulis, and S.~A. Tretyakov,
  ``{Intelligent Metasurfaces with Continuously Tunable Local Surface Impedance
  for Multiple Reconfigurable Functions.}'' \emph{arXiv Applied Physics}, pp.
  1--8, 2018. [Online]. Available: \url{http://arxiv.org/abs/1811.10082}
\BIBentrySTDinterwordspacing

\bibitem{Srinivasan2004}
J.~Srinivasan, S.~Adve, P.~Bose, and J.~Rivers, ``{The impact of technology
  scaling on lifetime reliability},'' in \emph{Proceedings of the DSN '04},
  2004, pp. 177--186.

\bibitem{Wan2016}
X.~Wan, M.~Q. Qi, T.~Y. Chen, and T.~J. Cui, ``{Field-programmable beam
  reconfiguring based on digitally-controlled coding metasurface},''
  \emph{Scientific Reports}, vol.~6, p. 20663, 2016.

\bibitem{Huang2017}
C.~Huang, B.~Sun, W.~Pan, J.~Cui, X.~Wu, and X.~Luo, ``{Dynamical beam
  manipulation based on 2-bit digitally-controlled coding metasurface},''
  \emph{Scientific Reports}, vol.~7, no. January, pp. 1--8, 2017.

\bibitem{Tasolamprou2014}
A.~C. Tasolamprou, L.~Zhang, M.~Kafesaki, T.~Koschny, and C.~M. Soukoulis,
  ``{Experimentally excellent beaming in a two-layer dielectric structure},''
  \emph{Optics Express}, vol.~22, no.~19, p. 23147, 2014.

\bibitem{Tasolamprou2017}
A.~C. Tasolamprou, T.~Koschny, M.~Kafesaki, and C.~M. Soukoulis,
  ``{Near-Infrared and Optical Beam Steering and Frequency Splitting in
  Air-Holes-in-Silicon Inverse Photonic Crystals},'' \emph{ACS Photonics},
  vol.~4, no.~11, pp. 2782--2788, 2017.

\bibitem{Kouvaros2018}
P.~Kouvaros, D.~Kouzapas, A.~Philippou, J.~Georgiou, L.~Petrou, and
  A.~Pitsillides, ``{Formal Verification of a Programmable Hypersurface},'' in
  \emph{Proceedings of the FMICS '18}, 2018, pp. 83--97.

\bibitem{BalanisBOOK}
C.~A. Balanis, \emph{{Antenna Theory: Analysis and Design}}, 3rd~ed., Wiley,
  Ed., 2005.

\bibitem{Zhang2018a}
L.~Zhang, R.~Y. Wu, G.~D. Bai, H.~T. Wu, Q.~Ma, X.~Q. Chen, and T.~J. Cui,
  ``{Transmission-Reflection-Integrated Multifunctional Coding Metasurface for
  Full-Space Controls of Electromagnetic Waves},'' \emph{Advanced Functional
  Materials}, vol.~28, no.~33, pp. 1--9, 2018.

\bibitem{Moccia2018}
M.~Moccia, C.~Koral, G.~P. Papari, S.~Liu, L.~Zhang, R.~Y. Wu, G.~Castaldi,
  T.~J. Cui, V.~Galdi, and A.~Andreone, ``{Suboptimal Coding Metasurfaces for
  Terahertz Diffuse Scattering},'' \emph{Scientific Reports}, vol.~8, no.~1,
  pp. 2--10, 2018.

\bibitem{Xu2017b}
H.-X. Xu, S.~Ma, X.~Ling, X.-K. Zhang, S.~Tang, T.~Cai, S.~Sun, Q.~He, and
  L.~Zhou, ``{Deterministic Approach to Achieve Broadband
  Polarization-Independent Diffusive Scatterings Based on Metasurfaces},''
  \emph{ACS Photonics}, vol.~5, pp. 1691--1702, 2017.

\end{thebibliography}
\end{document}